# Optical tuning and ultrafast dynamics of high-temperature superconducting terahertz metamaterials

Ranjan Singh[1], Jie Xiong[1,2], Abul K. Azad[1], Hao Yang[1,3], Stuart A. Trugman[1], Q. X. Jia[1], Antoinette J. Taylor[1], and Hou-Tong Chen[1*]

[1] Center for Integrated Nanotechnologies, Los Alamos National Laboratory, Los Alamos, New Mexico 87545, USA

[2] State Key Lab of Electronic Thin Films and Integrated Devices, University of Electronic Science and Technology of China, Chengdu 610054, China

[3] Jiangsu Key Laboratory of Thin Films, School of Physical Science and Technology, Soochow University, Suzhou 215006, China

[*] Email: chenht@lanl.gov

**Abstract**

Through the integration of semiconductors or complex oxides into metal resonators, tunable metamaterials have been achieved by a change of environment using an external stimulus. Metals provide high conductivity to realize a strong resonant response in metamaterials; however, they contribute very little to the tunability. The complex conductivity in high-temperature superconducting films is highly sensitive to external perturbations, which provides new opportunities in achieving tunable metamaterials resulting directly from the resonant elements. Here we demonstrate ultrafast dynamical tuning of resonance in the terahertz (THz) frequency range in $YBa_2Cu_3O_{7-\delta}$ (YBCO) split-ring resonator arrays excited by near-infrared femtosecond laser pulses. The photoexcitation breaks the superconducting Cooper pairs to create the quasiparticle state. This dramatically modifies the imaginary part of the complex conductivity and consequently the metamaterial resonance in an ultrafast timescale. We observed resonance switching accompanied with a wide range frequency tuning as a function of photoexcitation fluence, which also strongly depend on the nano-scale thickness of the superconducting films. All of our experimental results are well reproduced through calculations using an analytical model, which takes into account the SRR resistance and kinetic inductance contributed from the complex conductivity of YBCO films. The theoretical calculations reveal that the increasing SRR resistance upon increasing photoexcitation fluence is responsible for the reduction of resonance strength, and both the resistance and kinetic inductance contribute to the tuning of resonance frequency.

**Introduction**

An important issue in resonant electromagnetic metamaterials is how to overcome the loss, particularly in the optical frequency range where the Drude response provides limited conductivity in metals [1-3]. However, one may also take advantage of the typically undesirable loss to realize certain functionalities such as switching or modulation of electromagnetic waves [4], where the metamaterial resonant response plays a critical role in enhancing such functionalities. There have been quite a few approaches in realizing tunability in hybrid metamaterials where materials such as semiconductors and complex oxides serve as the metamaterial substrate or spacer, or they are integrated at specific regions of metamaterial resonators [5-24]. The resonant response in hybrid metamaterials could be switched and/or frequency tuned when an external stimulus is applied such as temperature [5-8], voltage bias [9-14], and photoexcitation [15-24]. The optical approach has received the most intensive attention in metamaterial photonics due to its simplicity, capability of ultrafast tuning, and potential of wide range applications. Usually it is the change of environment of metamaterial resonators that results in the resonance tuning; however, metals that compose metamaterial resonators contribute very little except for providing high conductivity to realize a strong resonance.

It has been shown that metal conductivity as well as thickness can affect metamaterial resonance [25,26], though such tunability is through design and fabrication. Noble metals are still the choice for fabricating metamaterial structures because of their high conductivity, which, however, makes the resonance tuning rely on the integration of additional materials able to respond to external stimuli. The complex conductivity in high-temperature superconducting films is highly sensitive to external perturbations such as temperature, magnetic fields, optical excitation, and electrical current. In fact, there have been a few recent demonstrations of superconducting metamaterials, which are of particular interest in loss reduction and resonance tuning [27-37], the latter has mostly been achieved by change of temperature [27-33,36,37] and application of magnetic fields [29,34], two approaches that are uniquely suited to manipulation of superconducting properties. Additionally, due to the strong frequency dependence of the complex conductivity, as well as the energy band gap in superconductors, superconducting metamaterials are limited to frequencies up to the terahertz (THz) frequency range.

In this work, we accomplish ultrafast resonance tuning in planar high-temperature superconducting metamaterials through near-infrared femtosecond photoexcitation. It breaks the superconducting Cooper pairs to create the quasiparticle state, which dramatically modifies the imaginary part of the complex conductivity and consequently tunes the metamaterial resonance. We observed the resonance switching accompanied with a wide range frequency tuning as a function of photoexcitation fluence, which also strongly depend on the nano-scale thickness of the superconducting films. We experimentally investigated the dynamics of the metamaterial resonance for various pump fluences, and the results are compared to and reproduced by calculations using an analytical model [28,33], which takes into account both the SRR resistance and kinetic inductance contributed from the complex conductivity of the superconducting films.

**Experiments and Results**

Epitaxial $YBa_2Cu_3O_{7-\delta}$ (YBCO) high-temperature superconducting films with thicknesses of 50, 100, and 200 nm were prepared on 500-μm-thick (100) $LaAlO_3$ (LAO) substrates using pulsed laser deposition, with film area size of 10 mm × 10 mm. The transition temperature was measured to be $T_c$ = 90 K. Square arrays of electric split-ring resonators (SRRs) [38], with a unit cell microscopic image shown in the inset to Fig. 1, were fabricated from the YBCO films through conventional photolithographic methods and wet chemical etching using 0.01% $H_3PO_4$ solution for several minutes. Under normal incidence, the YBCO films and metamaterials were characterized at a temperature of 20 K using an optical-pump THz-probe (OPTP) spectrometer [16,39] incorporated with a continuous flow liquid helium cryostat. The output near-infrared laser beam (50 fs, 3.2 mJ/pulse at 800 nm with a 1 kHz repetition rate) was split into two parts with one being used for THz generation-detection and the other for optical excitation (pump) of the metamaterial samples. The pump beam has a beam diameter of ~1 cm, much larger than the focused THz spot diameter of ~3 mm at the sample, providing mostly uniform excitation over the YBCO films and metamaterials. Variation of pump-probe time delay was realized by using a motion stage to change the pump beam optical path. For various pump powers and pump-probe time delays, THz pulses transmitting through the YBCO films and metamaterials, as well as a blank LAO substrate as the reference, were measured in the time-domain, i.e., recording the time-varying electric field of the impulsive THz radiation. The transmission amplitude and phase spectra were directly obtained by performing fast Fourier transformation of the time-domain signal and normalized to that of the reference. The complex conductivity of the YBCO films was extracted from the measured transmission amplitude and phase spectra through the inversion of Fresnel equations.

The dynamics of bare YBCO films were first measured under near-infrared photoexcitation with various fluences. Figure 1 shows the transmitted THz peak signal through a 100-nm-thick YBCO film as a function of time delay between the near-infrared pump and THz probe pulses, for optical pump fluences of 0.05 and 0.3 $mJ/cm^2$, or powers of 50 and 300 mW, respectively. At low temperatures, the YBCO films are superconducting with high imaginary conductivity, which results in very low transmission of THz radiation. The near-infrared photons break the superfluid

Cooper pairs in the superconducting YBCO films and create the quasiparticle state. It dramatically reduces the imaginary conductivity and therefore increases the transmitted THz signal. When time delay increases, the transmitted THz signal decreases due to the recombination of quasiparticles to form Cooper pairs. Under low fluence excitation, the relaxation time is about 5 ps dominated by the recombination; however, it has a much longer tail for higher pump fluences due to significant thermal effects. The observed small peak in Fig. 1 at the time delay of 16 ps (which is ~7 ps after the main transmission peak) arises from reflected laser pulses at the back surface of the substrate. We have carried out dynamical measurements of the resonance of the YBCO metamaterials for various pump powers, and the results are very similar to those shown in Fig. 1 for the bare YBCO film.

The amplitude spectra of THz transmission through the 100-nm-thick metamaterial sample were then measured at 20 K for various photoexcitation fluences at time delays indicated by the arrows in Fig. 1. In the absence of a pump pulse, the metamaterial exhibits a strong resonant response, as shown by the red curves in Figs. 2A-2D with a resonant transmission minimum of 10% or power transmission of 1%, which is comparable to that in metamaterial samples fabricated from gold with the same thickness. This also facilitates the use of high-temperature superconductors in metamaterials replacing noble metals with additional tuning capability. Figure 2A shows the resonant transmission spectra for various pump powers when THz pulses pass through the metamaterial sample a few picoseconds before the optical pump pulses (time position I in Fig. 1). The small resulting change indicates that the excitation in the YBCO metamaterial has relaxed nearly completely during the time interval of 1 ms between the optical pump pulses. Immediately following the femtosecond photoexcitation (time position II), the metamaterial resonance is significantly weakened and red-shifted with increasing pump power, as shown in Fig. 2B. The resonance disappears at a pump power of 100 mW. Increasing the pump-probe time delay results in recovery of metamaterial resonance strength due to carrier relaxation, as revealed in Figs. 2C and 2D for positions III and IV, respectively, both of which are before the arrival of the reflected optical pulse from the back surface of the substrate. With the same pump power, the resonance strength increases and its frequency shifts back when the time delay increases. However, when the pump power is high, e.g. 500 mW, the thermal effects last much longer and the metamaterial resonance does not recover within the time delay of tens picoseconds in our measurements, which is consistent with the dynamics measurements shown in Fig. 1.

It has been shown that the resonance tuning in superconducting metamaterials through change of temperature is strongly dependent on the film thickness [33]. We also fabricated identical metamaterial structure from YBCO films with two other thicknesses of 50 and 200 nm. Measured at the pump-probe time delay position III, the pump power dependent resonant transmission spectra are shown in Figs. 2E and 2F. It reveals that increasing YBCO film thickness results in a higher resonance frequency, stronger resonance strength, smaller frequency tuning range, and requires higher pump power in tuning the resonance. In the absence of a pump pulse, the resonance frequency is 0.460, 0.553, and 0.612 THz, and the resonance transmission minimum is 0.28, 0.11, and 0.03, respectively, for the YBCO metamaterials with thickness of 50, 100, and 200 nm. For the 50-nm-thick YBCO metamaterial, 100 mW pump laser power is sufficient to completely switch off the resonance; while for the 200-nm-thick YBCO metamateria,l even 1000 mW pump laser power is insufficient to switch off the resonance.

**Discussion**

The complex conductivity plays an essential role in the resonance tuning in high-temperature superconducting metamaterials [33]. Figure 3 shows the complex conductivity at 0.55 THz of the 100-nm-thick YBCO film, measured at 20 K as a function of pump power at different pump-probe time delays. When there is no optical pump, the imaginary part is about one order of magnitude higher than the real part. For time position I, there is little change in the conductivity, shown in Fig. 3A, with increasing pump power, which confirms that the 1 ms time interval of the laser pulses is sufficient for nearly-complete thermal relaxation. Shown in Fig. 3B for the pump-probe time position II, the imaginary conductivity dramatically decreases by orders of magnitude as the superfluid Cooper pairs are broken into quasiparticles by the near infrared photons, while the real conductivity does not exhibit a significant change (at most a factor of 2) as it arises from the Drude response of quasiparticles. Shown in Figs. 3C and 3D for the pump-probe time positions III and IV, respectively, the imaginary conductivity increases with increasing pump-probe time delay, which allows for recovery of the superconductivity.

There have been a few discussions that surface resistance and kinetic inductance contribute to resonance tuning in superconducting metamaterials [28,33]. The surface impedance (in Ω/Square) of a superconducting film can be calculated from the measured complex conductivity [33]:

$$\tilde{Z}_S = R_S + iX_S = Z_0 \frac{n_3 + i\tilde{n}_2 \cot(\tilde{\beta}d)}{\tilde{n}_2^2 - n_3^2} \cong i\frac{Z_0}{\tilde{n}_2}\cot(\tilde{\beta}d), \qquad (1)$$

where the "~" indicates complex value, $Z_0 = 377\ \Omega$ is the vacuum impedance, $n_3 = 4.8$ is the LAO substrate refractive index, $\tilde{n}_2 = \sqrt{i\tilde{\sigma}/\varepsilon_0\omega}$ is the complex refractive index of the YBCO film, $\tilde{\sigma}$ is the measured complex conductivity, $\tilde{\beta} = \tilde{n}_2\omega/c_0$ is the complex propagation constant where $c_0$ is the light velocity in vacuum, and $d$ is the thickness of the superconducting film. The surface resistance $R_S$ and reactance $X_S = \omega L_S$ are shown in Fig. 4 as a function of pump power for the 100-nm-thick YBCO film, which are calculated at 0.55 THz and at different pump-probe time delays. The surface resistance $R_S$, which represents the loss, increases rapidly with pump power; while the surface reactance $X_S$ increases first with lower pump power, reaches a maximum value, and then drops with further increasing pump power. As we compare the results in Figs. 3 and 4, it is interesting to find that the maximum surface reactance, and therefore the maximal frequency tuning, occurs at a pump power where the real and imaginary parts of the complex conductivity are equal. However, this is not really surprising since it can be easily verified using Eq. (1).

At resonance, the SRR reactance is absent and the surface resistance of the SRR layer can be calculated from the YBCO surface resistance, $R = [(A - g)/w]R_S$, where $A = 64$ μm is the circumference of the SRR current loop, $g = 4$ μm is the split gap width, and $w = 4$ μm is the strip width of the SRR. By treating the SRR layer as a shunt resistor in a transmission line, the resonant transmission minimum can then be calculated for the pump-probe time positions I-IV, with the results shown in Figs. 5A-5D, respectively. Compared to the experimental results, which are also shown in Fig. 5, there is good agreement, i.e., both the experiments and calculations reveal an increase in resonant transmission minimum (decreasing resonance strength) with pump power.

The frequency of the fundamental resonance in SRRs is typically given by $\omega_0 = 1/\sqrt{LC}$, where $L$ and $C$ are the loop inductance and gap capacitance, respectively. However, when the ohmic loss is significant, the resistance $R$ in an equivalent circuit of SRR has to be considered [33]:

$$\omega_0^2 = \frac{1}{LC} - \frac{R^2}{4L^2}. \tag{2}$$

In superconducting SRRs, the inductance $L$ includes two contributions: Faraday inductance $L_F$ which is determined and can be estimated by the geometry and dimensions of the SRR, and kinetic inductance $L_k = [(A-g)/w](X_S/\omega)$ which is associated with the kinetic energy in superconducting charge carriers. The capacitance $C$ can be derived using Eq. (2) from simulated resonance frequency $\omega'_0$ when assuming perfect conducting SRR, i.e. $L = L_F$, $R = 0$, and therefore $C = (\omega_0'^2 L_F)^{-1}$. With all these derived parameters, we can calculate the resonance frequency of the YBCO metamaterial as a function of pump power, as shown in Figs. 5E-5H for the pump-probe time positions I-IV, respectively. It is again compared to the experimental results with good agreement. Here we have limited our discussion to 80 mW pump power because the significant thermal effects result in diminishing metamaterial resonance strength for further increasing pump power over 100 mW. Upon photoexcitation, both the resistance and inductance increase, and the overall effect is a reduction of resonance frequency. Increasing the pump-probe time delay recovers both the resonance strength and frequency due to the formation of Cooper pairs on an ultrafast timescale.

Finally, the above model can also well explain the thickness dependent YBCO metamaterial resonance and tuning shown Fig. 6. According to Eq. (1), reducing the film thickness $d$ results in a larger surface resistance $R_S$ and consequently smaller resonance strength, i.e., a larger resonant transmission minimum, as verified in Fig. 6A. The surface reactance $X_S$ increases with reduced film thickness $d$, causing a higher kinetic inductance $L_K$ and lower resonance frequency $\omega_0$. The higher value of kinetic inductance $L_K$ also results in a larger change upon photoexcitation, causing the wider tuning range of the resonance frequency, as shown in Fig. 6B. The different requirement of pump power in the resonance tuning for YBCO metamaterials of varying thickness may be resulted from the limited penetration depth of the pump laser.

## Conclusions

We have shown the ultrafast resonance tuning behavior in high-temperature superconducting THz metamaterials through femtosecond near-infrared photoexcitation. Increasing the pump power results in a reduced resonance strength and red-shifting of frequency. Under low fluence photoexcitation, the relaxation time is ~5 ps; further increasing the pump fluence also introduces significant thermal effects with much longer relaxation time. The dynamics of the metamaterial resonance follows that of the superconducting films. Using experimentally measured complex conductivity of the superconducting films, we are able to reproduce such behavior through calculations by including the ohmic resistance and kinetic inductance. The theoretical calculations reveal that the increasing SRR resistance upon increasing photoexcitation fluence is responsible for the reduction of the resonance strength, and both the resistance and kinetic inductance contribute to the tuning of the resonance frequency. We have also shown that the nano-scale thickness of the superconducting films plays an important role in the frequency tuning range and requirement of optical pump fluence. Thinner superconducting metamaterials require lower pump fluence to switch the resonance with a larger frequency tuning range. While the demonstrated metamaterial resonance tuning is not practical at optical frequencies due to the existence of the superconducting energy gap, such ultrafast dynamical resonance tuning would be useful in realizing multifunctional microwave and THz metamaterial and plasmonic devices by further design and materials integration.

## Acknowledgements

We acknowledge support from the Los Alamos National Laboratory LDRD Program. This work was performed, in part, at the Center for Integrated Nanotechnologies, a US Department of Energy, Office of Basic Energy Sciences Nanoscale Science Research Centre operated jointly by Los Alamos and Sandia National Laboratories. Los Alamos National Laboratory, an affirmative action/equal opportunity employer, is operated by Los Alamos National Security, LLC, for the National Nuclear Security Administration of the US Department of Energy under contract DE-AC52-06NA25396.

**Figure Captions**

**Fig. 1 Dynamics of the 100-nm-thick bare YBCO film upon femtosecond near-infrared photoexcitation.** The transmitted THz peak signal was measured as a function of pump-probe time delay at 20 K for 50 and 300 mW photoexcitation powers. Position I indicates that THz probe pulses fail to transmit through the sample before the photoexcitation, and positions II, III and IV are at different pump-probe time delays where YBCO metamaterials are measured. Inset: a microscopic image of one unit cell of the 100-nm-thick YBCO metamaterial.

**Fig. 2 THz transmission amplitude spectra of YBCO metamaterials for various pump powers.** The spectra shown in **A-D** are for the 100-nm-thick YBCO metamaterial and the respective pump-probe time positions I-IV. **E** and **F,** THz transmission amplitude spectra for YBCO metamaterials with 50 and 200 nm thicknesses, respectively, measured at the pump-probe time position III.

**Fig. 3 Experimentally measured complex conductivity as a function of pump power for the 100-nm-thick bare YBCO film. A-D**, Real and imaginary parts of the complex conductivity are retrieved at 0.55 THz and measured at 20 K, for pump-probe time positions I-IV, respectively.

**Fig. 4 Calculated surface impedance as a function of pump power for the 100-nm-thick bare YBCO film. A-D**, The surface resistance and reactance are calculated using the experimental conductivity shown in Fig. 3 for the respective pump-probe time positions I-IV.

**Fig. 5 Experimentally measured and theoretically calculated transmission minimum and resonance frequency. A-D**, Resonant transmission minimum, and **E-H**, resonance frequency as a function of pump power for the 100-nm-thick YBCO metamaterial sample at the pump-probe time positions I-IV, respectively.

**Fig. 6 Experimentally measured resonance tuning for YBCO metamaterials of different thickness. A**, Resonant transmission minimum, and **B**, resonance frequency as a function of pump power for the YBCO metamaterials of 50, 100, 200 nm thicknesses, measured at the pump-probe time position III.

Figure 1

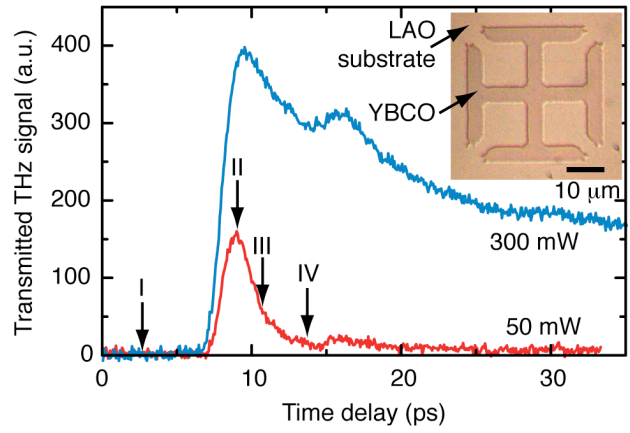

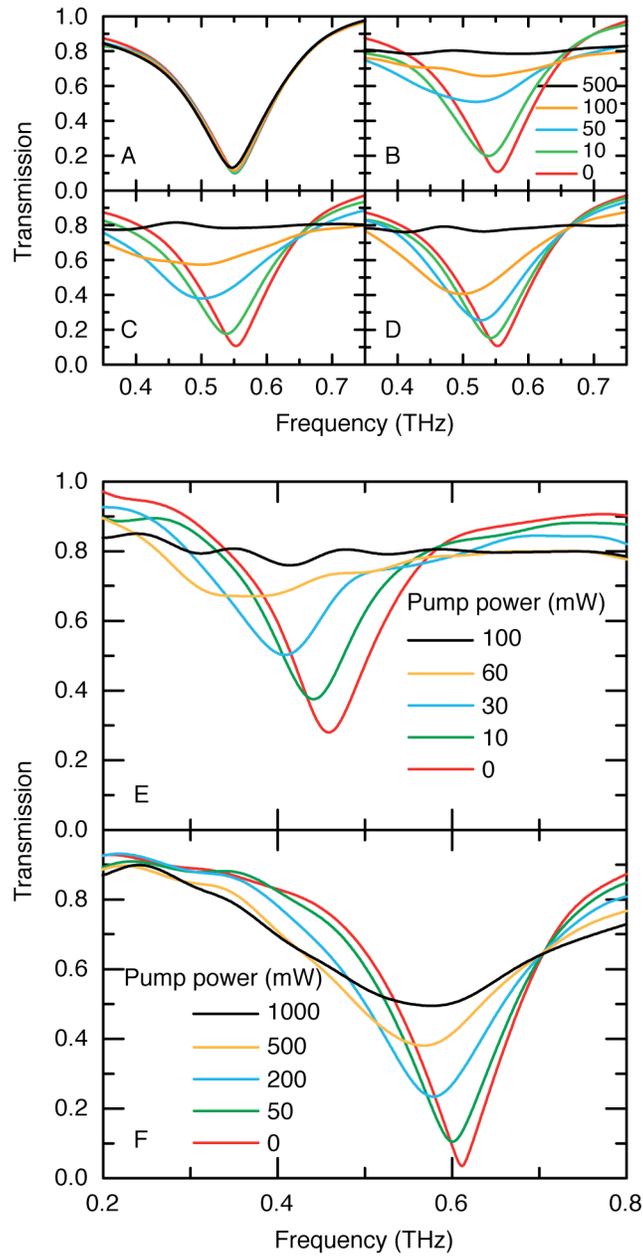

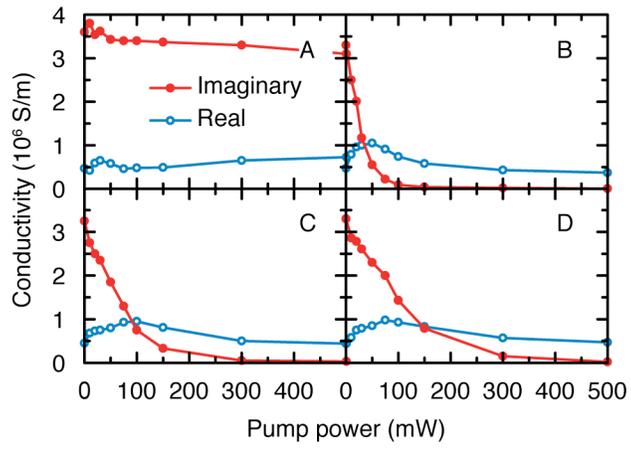

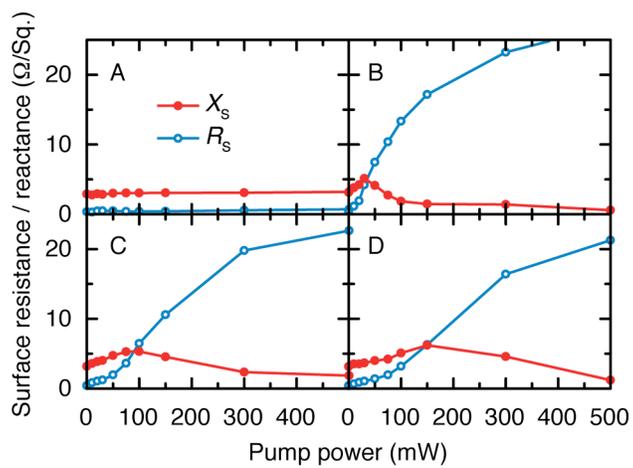

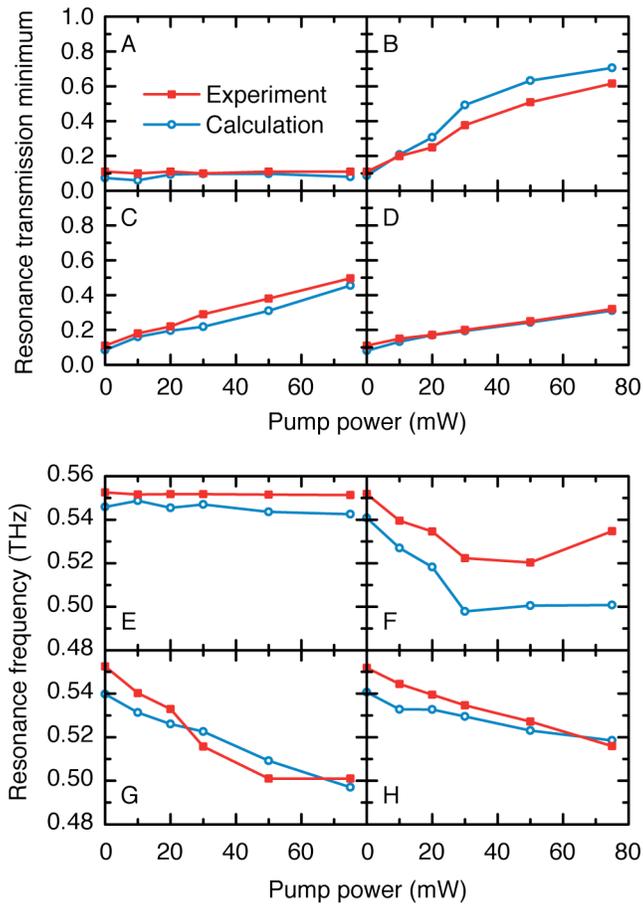

Figure 6

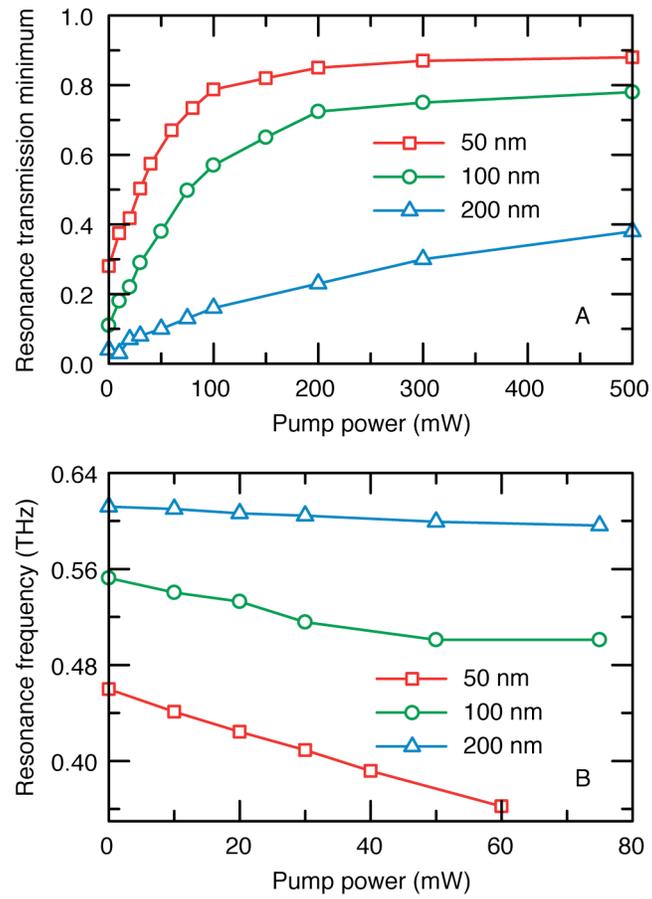